

\input phyzzx


\def\UCR{\address{University of California at Riverside\break
                  Department of Physics\break
                  Riverside, California 92521--0413; U.S.A. \break
                  WUDKA{\rm @}UCRPHYS}}

%
%
%
\def\and{{\it\&}}

\def\gesim{\,{\raise-3pt\hbox{$\sim$}}\!\!\!\!\!{\raise2pt\hbox{$>$}}\,}
\def\lesim{\,{\raise-3pt\hbox{$\sim$}}\!\!\!\!\!{\raise2pt\hbox{$<$}}\,}
\def\boldoverdot{\,{\raise6pt\hbox{\bf.}\!\!\!\!\>}}

\def\ie{{\it i.e.}\ }

\def\ibid{{\it ibid.}\ }
\def\etal{{\it et. al.}\ }

\def\lcal{{\cal L}}

\def\ocal{{\cal O}}
\def\pcal{{\cal P}}

\def\lBB{ \hbox{{\smallii I}}\!\hbox{{\smallii L}} }

\def\diag{\hbox{\diag}}
\def\sm{Standard Model}

\def\gev{\hbox{GeV}}
\def\tev{\hbox{TeV}}

%
%
%

%

%
\def\inbox#1{\vbox{\hrule\hbox{\vrule\kern5pt
     \vbox{\kern5pt#1\kern5pt}\kern5pt\vrule}\hrule}}
\def\sqr#1#2{{\vcenter{\hrule height.#2pt
      \hbox{\vrule width.#2pt height#1pt \kern#1pt
         \vrule width.#2pt}
      \hrule height.#2pt}}}

\def\today{\ifcase\month\or
  January\or February\or March\or April\or May\or June\or
  July\or August\or September\or October\or November\or December\fi
  \space\number\day, \number\year}
\def\pmb#1{\setbox0=\hbox{#1}%
  \kern-.025em\copy0\kern-\wd0
  \kern.05em\copy0\kern-\wd0
  \kern-.025em\raise.0433em\box0 }
\def\up#1{^{\left( #1 \right) }}
\def\lowti#1{_{{\rm #1 }}}
\def\inv#1{{1\over#1}}
\def\su#1{{SU(#1)}}

\def\antes{}
\def\despues{.}
%

%
\def\sumprime_#1{\setbox0=\hbox{$\scriptstyle{#1}$}
  \setbox2=\hbox{$\displaystyle{\sum}$}
  \setbox4=\hbox{${}'\mathsurround=0pt$}
  \dimen0=.5\wd0 \advance\dimen0 by-.5\wd2
  \ifdim\dimen0>0pt
  \ifdim\dimen0>\wd4 \kern\wd4 \else\kern\dimen0\fi\fi
\mathop{{\sum}'}_{\kern-\wd4 #1}}
\def\sumbiprime_#1{\setbox0=\hbox{$\scriptstyle{#1}$}
  \setbox2=\hbox{$\displaystyle{\sum}$}
  \setbox4=\hbox{${}'\mathsurround=0pt$}
  \dimen0=.5\wd0 \advance\dimen0 by-.5\wd2
  \ifdim\dimen0>0pt
  \ifdim\dimen0>\wd4 \kern\wd4 \else\kern\dimen0\fi\fi
\mathop{{\sum}''}_{\kern-\wd4 #1}}
\def\sumtriprime_#1{\setbox0=\hbox{$\scriptstyle{#1}$}
  \setbox2=\hbox{$\displaystyle{\sum}$}
  \setbox4=\hbox{${}'\mathsurround=0pt$}
  \dimen0=.5\wd0 \advance\dimen0 by-.5\wd2
  \ifdim\dimen0>0pt
  \ifdim\dimen0>\wd4 \kern\wd4 \else\kern\dimen0\fi\fi
\mathop{{\sum}'''}_{\kern-\wd4 #1}}
%
%
\newcount\chapnum
\def\clearchap{\chapnum=0}
\def\chap#1{\clearsect\clearprob
\global\advance\chapnum by 1 \par\vskip .5 in\par%
\centerline{{\bigboldiii\antes\the\chapnum\despues\ #1}}}
\newcount\sectnum
\def\clearsect{\sectnum=0}
\def\sect#1{\clearprob\global\advance\sectnum by 1 \par\vskip .25 in\par%
\noindent{\bigboldii\the\chapnum.\the\sectnum:\ #1}\nobreak}
\newcount\yesnonum
\def\clearyesno{\yesnonum=0}
\def\verify{\global\advance\yesnonum by 1{\bigboldi (VERIFY!!)}}
\def\tocheck{\par\vskip 1 in{\bigboldv TO VERIFY: \the\yesnonum\ ITEMS.}}
\newcount\notenum

\def\note#1{\global\advance\notenum by 1{ \bf $<<$ #1 $>>$ } }
\def\noteout{\par\vskip 1 in{\bigboldiv NOTES: \the\notenum.}}
\newcount\borrownum

\def\borrow{\global\advance\borrownum by 1{\bigboldi BORROWED BY:\ }}
\def\borrowed{\par\vskip 0.5 in{\bigboldii BOOKS OUT:\ \the\borrownum.}}
\newcount\refnum

\def\ref#1{\global\advance\refnum by 1\item{\the\refnum.\ }#1}
\def\stariref#1{\global\advance\refnum by 1\item{%
               {\bigboldiv *}\the\refnum.\ }#1}
\def\stariiref#1{\global\advance\refnum by 1\item{%
               {\bigboldiv **}\the\refnum.\ }#1}
\def\stariiiref#1{\global\advance\refnum by 1\item{
               {\bigboldiv ***}\the\refnum.\ }#1}
\newcount\probnum
\def\clearprob{\probnum=0}
\def\prob{\global\advance\probnum by 1 {\medskip $\triangleright$\
\undertext{{\sl Problem}}\ \the\chapnum.\the\sectnum.\the\probnum.\ }}
\newcount\probchapnum

\def\probchap{\global\advance\probchapnum by 1 {\medskip $\triangleright$\
\undertext{{\sl Problem}}\ \the\chapnum.\the\probchapnum.\ }}
\def\undertext#1{$\underline{\smash{\hbox{#1}}}$}
%
%
%

\def\UCR{
{{\it University of California at Riverside\break
                  Department of Physics\break
                  Riverside, California 92521--0413; U.S.A. \break
                  WUDKA{\rm @}UCRPHYS}}}

%
%
%
%
\font\sanser=cmssq8

%

%

%
\font\bigboldi=cmbx10 scaled\magstep1
\font\bigboldii=cmbx10 scaled\magstep2
\font\bigboldiii=cmbx10 scaled\magstep3
\font\bigboldiv=cmbx10 scaled\magstep4
\font\bigboldv=cmbx10 scaled\magstep5
\font\small=cmr8
\font\smalli=cmr8 scaled\magstep1
\font\smallii=cmr8 scaled\magstep2

%

%

%
%
%
%
\clearchap
\clearyesno
\headline={\ifnum\pageno>0   {\smalli \title (\today)} \hfil {\small Page
\folio } \else\hfil\fi}
\newdimen\fullhsize
\newdimen\fullvsize
\newbox\leftcolumn
\def\fullline{\hbox to\fullhsize}
\gdef\twocol{\fullhsize=9.75in
\hsize=4.6in
\vsize=7in
\advance\hoffset by -.5 in
\def\makeheadline{\vbox to 0pt{\vskip-.4in
  \fullline{\vbox to8.5pt{}\the\headline}\vss}
   \nointerlineskip}
\def\makefootline{\baselineskip=24pt
    \fullline{\the\footline}}
\let\lr=L
\output{\if L\lr
   \global\setbox\leftcolumn=\columnbox \global\let\lr=R
  \else \doubleformat \global\let\lr=L\fi
        \ifnum\outputpenalty>-2000 \else\dosupereject\fi}
\def\doubleformat{\shipout\vbox{\makeheadline
     \fullline{\box\leftcolumn\hfil\columnbox}\makefootline
     }\advancepageno}
\def\columnbox{\leftline{\pagebody}}
\nopagenumbers
\hfuzz=3pt}

\def\title{Paper on $e_R p \rightarrow \nu X $ at HERA}

\def\pol{\hbox{$\pcal$}}
\def\lt{\tilde \Lambda }
\def\lum{\hbox{$\lBB$}}
\def\mnu{{m_\nu}}
\def\mnus{ {\mnu^2 \over s } }
\def\sm{Standard Model}
\def\nbck{N \lowti{ bcgd. } }

\def\theabstract{We study the sensitivity of HERA to new physics using the
helicity suppressed reaction $e_R p \rightarrow \nu X $, where the final
neutrino can be a standard model one or a heavy neutrino. The approach is model
independent and is based on an effective lagrangian parametrization. It is
shown
that HERA will put significant bounds on the scale of new physics, though,
in general, these are more modest than previously thought. If
deviations from the standard model are observed in the above processes, future
colliders such as the SSC and LHC will be able to directly probe the physics
responsible for these discrepancies}

\unnumberedchapters
\PHYSREV
\pubnum{113}
\date{\today}
{\titlepage
\vskip -.2 in
\title{ {\bigboldiii Anomalous neutrino reactions at HERA }}
\doublespace
\author{{ Jos\'e Wudka }}
\address{\UCR}
\abstract
\doublespace
\theabstract
\endpage}
\doublespace

\REF\polar{
H.F. Contopanagos, M.B. Einhorn, {\sl Nucl. Phys.} {\bf B377}, 20, (1992).}
{\bf 1.} The fact that HERA will provide a polarized electron beam opens the
possibility of performing experiments which are sensitive to physics beyond the
standard model [\polar]. In this report the reaction for which a right handed
electron scatters of a proton creating a neutrino will be studied, both for the
standard model particle spectrum and when a right-handed massive neutrino is
added to it. This reaction is enormously suppressed in the standard
model\foot{{\sanser The \sm\ charged currents are left handed and a chirality
flip must occur, thus the corresponding cross section is proportional to $ (
m_e / E_e )^2 \sim 4 \times 10^{ - 10 } $.}} and is therefore a good process
in which to
look for new physics. The approach which we will use consists in parametrizing
the effects of new interactions using an effective lagrangian. This has the
advantage of preserving all symmetries of the \sm\ and is completely model
independent with respect to the underlying physics. The approach also provides
reliable estimates for the magnitudes of the coefficients of the effective
operators.

\REF\blv{W. B\"uchmuler, \etal {\sl Phys. Lett.} {\bf 197B}, 379 (1987).}
\REF\curcur{
E.J. Eichten, \etal {\sl Phys. Rev. Lett. }, {\bf50}, 811 (1983).
R. R\"uckl, {\sl Phys. Lett.}, {\bf129B}, 363 (1983);
            {\sl Nucl. Phys.}, {\bf B234}, 91 (1984).
R.J. Cashmore, \etal {\sl Phys. Rep.}, {\bf122}, 275 (1985),
and references therein.}
The most important operators responsible for left handed neutrino production
were identified long ago in Ref.\blv\ where restrictions on these interactions
stemming from low energy meson physics were obtained, and where the effects on
the Callan--Gross relation were determined. In this report we will present a
parallel study where the effects of these operators on polarized cross sections
are considered. \foot{{\sanser These operators must violate chirality and are
therefore {\it not} of the type current$\times$current which have been
extensively studied [\curcur].}}

\REF\bs{C.J.C Burges and H.J.Schnitzer,
                {\sl Phys. Lett.}, {\bf 134B}, 329 (1984).}
\REF\prev{M.A. Doncheski, {\sl Z. Phys.}, {\bf C52}, 527, (1991).}
The processes under consideration has been identified before in
[\bs,\prev]; and in [\prev] it is claimed that HERA will be sensitive
to new physics up to a scale of a few \tev. We will see that this
result is strongly dependent on the assumptions regarding the physics
underlying the standard model: if it is weakly coupled (such situation
is realized, for example by supersymmetry), sensitivity to the scale
of new physics, which we call $ \Lambda $, is much weaker. (It must be
emphasized that the meaning of $ \Lambda $ is very clear within this
approach: it represents the mass of the lightest excitation in the
underlying theory.)

\REF\bw{
C.J.C Burges and H.J.Schnitzer,
                 {\sl Nucl. Phys.}, {\bf B228}, 464 (1983).
C.N. Leung, \etal {\sl Z. Phys.}, {\bf C31}, 433 (1986).
W. Buchm\"uller and D. Wyler, {\sl Nucl. Phys.}, {\bf B268}, 621 (1986).}
\bigskip {\bf 2.} We consider first the reaction $ e_R p \rightarrow \nu_L X $,
where $p$ represents a proton and $ \nu_L $ the electron's
neutrino\foot{{\sanser We will ignore inter-generation mixing as the scale for
the corresponding operators is presumably much larger than $ \Lambda $, see
Refs. \bw.}}. As mentioned above, the \sm\ cross section for this process is
essentially zero; in contrast, physics underlying the \sm\ can generate
effective interactions which contribute significantly to this reaction. The
relevant operators are of two types and can be found in the complete
compilation of B\"uchmuller and Wyler [\bw] (for an operator not included in
this list see Ref. \blv)

The contributing four fermi operators are $$ \eqalign{ \ocal_{ \ell q
} &= (
\bar \ell e ) \epsilon ( \bar q u ) \cr \ocal_{ q d e } &= ( \bar \ell e ) (
\bar d q ) \cr \ocal_{ \ell q } ' & = ( \bar \ell u ) \epsilon ( \bar q e ) \cr
} \eqn \theops $$ where we adopted the notation and conventions of B\"ucmuller
and Wyler in Refs. \bw: $
\ell $ and $q$ denote the left handed lepton and quark doublets respectively, $
e, \ u $ and $d$ denote the right handed electron, up and down quark fields,
and $ \epsilon = i \sigma_2 $.

The remaining operators containing \sm\ fields which contribute to the process
at hand are $$ \eqalign{ \ocal_{ D e } &= ( \bar \ell D_\mu e ) ( D^\mu \phi )
\cr \ocal_{ \bar D e } &= ( \overline{ D_\mu \ell } e ) ( D^\mu \phi ) \cr
\ocal_{ e W } &= g \bar \ell \sigma^{ \mu \nu } \tau^I e \phi W^I_{ \mu \nu }
\cr } \eqn\moreops $$ where $ \phi $ denotes the scalar doublet, $ D_\mu $ the
covariant derivative, $ \tau^I $ the usual Pauli matrices, and $ W^I_{ \mu \nu
}$ the $ \su2 _L $ gauge field strength tensor with $g$ the corresponding
coupling constant. These operators contribute via $t$ channel $W$ exchange.

The lagrangian is therefore $$ \lcal_L = \lcal\lowti{St. Model} + \inv{
\Lambda^2 } \sum_i \lambda_i \ocal_i \eqn\leftl $$ (the sum over $i$ runs over
the above six terms, the subscript $L$ refers to the neutrino's helicity). We
have kept only the operators of lowest dimension contributing at tree level, as
they give the leading contributions.

\REF\ew{M.B. Einhorn and J. Wudka, in preparation.}
To estimate the magnitude of the $ \lambda_i $ we note first that, the
operators \theops\ can be generated at tree level by the underlying particles,
while \moreops\ can appear only at one loop [\ew]. This, together with the
assumption that the underlying physics is weakly coupled, implies that $
\lambda_i \lesim 1 $ for \theops, while $ \lambda_i \lesim 1/ 16 \pi^2 $ for
\moreops. The $ \lambda $'s will also contain some coupling constants from the
underlying theory, we will assume that these are of the same order as the weak
gauge coupling, then $ \lambda_i \sim 0.44 $ for \theops, and $ \lambda_i \sim
2 \times 10^{ - 3 } $ for \moreops. With these estimates $ \Lambda $ will
correspond to the mass of a low lying excitation in the underlying physics.

There are some experimental constraints on the coefficients $ \lambda_i $: from
$K$ and $ \pi $ decays it is known that [\blv] $$ \lambda_{ q d e } \simeq 0
\qquad \hbox{and} \qquad \lambda'_{ \ell q } \simeq 2 \lambda_{ \ell q }
\eqn\constr $$ which we will henceforth adopt. Within the framework of this
work these conditions are assumed to be the result of some (unknown) constraint
stemming from the underlying physics.

While we have assumed that the physics at scale $ \Lambda $ will generate all
operators in \theops\ and \moreops, this may not to be the case: one could
imagine situations where \theops\ are generated at a scale $ \Lambda $ while
\moreops\ are produced at a different scale $ \Lambda' \ll \Lambda
$\foot{{\sanser For the reaction under study the contributions from \theops\
and \moreops\ are comparable when $ \Lambda \sim 25 \Lambda' $.}}; if this is
the case HERA will be insensitive to the physics responsible for \theops\ and
\moreops\ for any interesting (\ie\ $ \gesim 100 \gev $) value of $ \Lambda $.

\REF\ehlq{
E. Eichten, \etal {\sl Rev. Mod. Phys.}, {\bf56}, 579 (1984);
                                   \ibid. {\bf58}, 1065 (1986).}
With \leftl\ we can calculate the cross section for the process at hand. Note
that the two types of operators \theops\ and \moreops\ will not interfere due
to helicity conservation at the quark vertex (provided we ignore quark masses,
as we will). The result is $$ { d \sigma_L \over dx dy } = { s \over 32 \pi \lt
^4 } x \left[ (2 - 3 y )^2 U + ( 2 - y )^2 \bar D \right] + { g^4 v^2 \over 64
\pi \Lambda_2^4 } \left[ { x^2 y ( 1 - y ) \over ( x y + m_w^2/s )^2 } ( U +
\bar D ) \right] \eqn\sigl $$ where $x$ and $y$ are the usual scaling
variables, $ v = \sqrt{2} \langle \phi \rangle \simeq 250 \gev $, $U$ and $
\bar D $ are the ($x$ and $y$ dependent) quark distribution functions; we also
defined $$ \lt = \Lambda / | \lambda_{ \ell q } | ^{ 1/2 } , \qquad \hbox{ and
} \qquad \lt_1 = \Lambda / | \lambda_{ De } - \lambda_{ \bar D e } + 8
\lambda_{ e W } | ^{ 1/2 } \eqn\eq . $$ The result of a numerical integration
using the EHLQ type II distribution functions [\ehlq] is (for $ \sqrt{ s } =
292 \gev $ with no cuts imposed) $$ \inv s \sigma_L = 0.0021 { 1 \over\lt^4 } +
0.00013 { 1 \over \lt_1^4 } \eqn\sigl $$ From the previous discussion we expect
$ \lt \sim 1.5 \Lambda $ and $ \lt \ll \lt_1 $, we will therefore ignore the
contributions from the operators \moreops.

Using \sigl\ we can estimate the sensitivity of HERA to the scale $ \Lambda $:
the above cross section will generate 15 events per year at HERA provided $
\Lambda \lesim 315 \gev $.

This result will be weakened for polarizations below 92\%; in this
case the statistical significance of the right-handed electron signal
must be considered. Suppose that we have a beam of polarization $ \pol
$, where $ \pol = 1 $ implies pure right handed electrons. Let $
\sigma_{ SM } $ be the \sm\ contribution to $ e_L p \rightarrow \nu_L
X $ via $t$-channel $W$ exchange; a simple calculation shows that
\foot{{\sanser There is a large class of operators which also modify
the couplings of the $W$ to the left handed weak currents, as well as
shifting the $W$ mass from it's standard model value [\bw]. We have
not included these contributions in $ \sigma_{ SM } $ since they
represent but small corrections and we are interested only in a
sensitivity estimate for $ \Lambda $.}}
$$ \sigma_{ SM } = { g^4 \over 8 \pi s } \int_0^1
dx dy { x U + x ( 1 - y ) ^2 \bar D \over \left( xy + m_W^2/s \right)^2 }
\simeq5.67 \times 10^{ - 7 } \gev^{ - 2 } , \quad ( \sqrt{s} = 314 \gev )
\eqn\eq $$ The number of signal events is $ N \lowti{ signal } = \pol \sigma_L
\lum $, where $ \lum $ is the luminosity ($ \lum = 4 \times 10^9 \gev^2 /$year
for HERA); the number of background events is $ \nbck = ( 1- \pol ) \sigma_{ SM
} \lum $; the condition for the signal to be statistically significant is $$ N
\lowti{ signal} > \sqrt{ \nbck + N \lowti{ signal } } . \eqn\statsign $$ which
determines the sensitivity to $ \Lambda $ given $ \pol $, the corresponding
graph is presented in \FIG\xfii{Sensitivity to $ \Lambda $ in the reaction $
e_R p \rightarrow \nu_L X $ as a function of the polarization \pol, assuming $
\left| \lambda_{ \ell q } \right| = 0.44 $. For $ \pol \le 0.9 $ the curve is
generated by \statsign, for $ \pol > 0.9 $ the curve corresponds to 15 events
per year at HERA.} figure \xfii. Note that for realistic values of \pol\ the
process is no longer rate dominated: once the signal is statistically
significant there will be more that 15 events per year.

\REF\pdb{K. Hikasa, \etal (Particle Data Book), {\sl Phys Rev.},
{\bf D45}, 1 (1992)}
\REF\leftright{
N.G. Deshpande \etal {\sl Phys. Rev.}, {\bf D44}, 837 (1991).
J. Gluza and M. Zra\l ek, {\sl Phys. Rev.}, {\bf D45}, 1693 (1992).}
\bigskip {\bf 3.} The same approach can be used to investigate the sensitivity
to $ \Lambda $ when a massive neutrino is included in the low energy spectrum.
In this case the characteristics of the above reaction (large missing
transverse momentum) can be also realized by the process $ e_R p \rightarrow
\nu_R X$ where $ \nu_R $ represents the right-handed polarized massive
neutrino. We will not deal here with the problem of including $ \nu_R $ in a
consistent model, we merely note that if its left handed partner (if present)
couples as an ordinary neutrino to the $W$ and $Z$ bosons, collider experiments
require the heavy neutrino to have a mass $ \mnu > 50 \gev $ [\pdb]. It must
also be mentioned that this type of object, if it exists, is probably not the
right handed partner of a known neutrino, as the corresponding models which
satisfy all constraints of flavor changing neutral currents
and neutrinoless double-beta decay, together with a
reasonable leptonic phenomenology, are difficult to construct without endowing
the right handed neutrinos with an enormous mass [\leftright]. Here we will
take a purely phenomenological approach adding such a particle to the \sm\
without reference to any specific model. We will assume that this massive
neutrino lies in the low energy spectrum of the model, but that its
interactions are generated by the physics underlying the \sm. Aside from a mass
term of the type $ \bar \ell \nu_R \phi $ there are two independent dimension
six operators which contribute to the reaction under consideration: $$
\eqalign{ \ocal_R \up 1 &= ( \bar d \gamma^\mu u ) ( \bar \nu_R \gamma_\mu e )
\cr \ocal_R \up 2 &= i (\phi^T \epsilon D_\mu \phi ) ( \bar \nu_R \gamma^\mu e
) . \cr } \eqn\nurops $$ As in the case for the left handed neutrinos, the
above two operators do not interfere due do helicity conservation at the quark
vertex (we again assume massless quarks) irrespective of the neutrino mass. The
cross section obtained from the lagrangian $$ \lcal_R = \lcal\lowti{St. Model}
+ \inv{ \Lambda^2 } \sum_{ i = 1 , 2 } \lambda_R \up i \ocal_R \up i \eqn\eq $$
is $$ \eqalign{ & {d \sigma_R \over dx dy } = { s \over 8 \pi \Lambda_{ R1 }^4
} \left\{ \left( x - \mnus \right) U + ( 1 - y ) \left[ x ( 1 - y ) + \mnus
\right] \bar D \right\} \cr & +{ s \over 8 \pi \Lambda_{ R2 }^4 } \left\{
\left( x - \mnus \right) \bar D + ( 1 - y ) \left[ x ( 1 - y ) + \mnus \right]
U \right\} \left( { m_W^2 \over x y s + m_W^2 } \right)^2 \cr } \eqn\dsigr $$
where $$ \Lambda^2_{ R1, R2 } = \Lambda^2 / | \lambda_R \up { 1 , 2 } | \eqn\eq
$$ and where $ x \ge \mnu^2/s $. As before we take $ \lambda_R \up 1 \sim 0.44$
and expect (with the previously mentioned caveats) $ \lambda_R \up 2 \ll
\lambda_R \up 1 $ whence the contributions from $ \ocal_R \up 2 $ will be
ignored.

If $ \nu_R $ exists, then a given experiment will not differentiate between the
reactions $ e_R p \rightarrow \nu_{ L , R } X $, and will actually measure the
``total'' cross section $ \sigma = \sigma_L + \sigma_R $. Using this we can
obtain a sensitivity plot for $ \Lambda $ by requiring the signal to be
statistically significant; as mentioned previously, this is the relevant
condition for polarizations below 92\%. Applying \statsign\ to the sum of the
above contributions (\ie\ using $ N \lowti{ signal } = \pol ( \sigma_R +
\sigma_L ) \lum $) we obtain the expected sensitivity to $ \Lambda $ as a
function of the right handed neutrino mass and the polarization. The results
are presented in \FIG\xfiii{Sensitivity to $ \Lambda $ (assuming $ \left|
\lambda_{ \ell q } \right| = \left| \lambda \up 1 _R \right| = 0.44 $) for 80\%
(dashes), 70\% (solid) and 60\% (dots) polarization, as a function of the right
handed neutrino mass. This includes the contributions from both the $ \nu_L $
and $ \nu_R $ processes.} figure \xfiii.

The effects of the massive neutrino are significant for $ \mnu \lesim 130 \gev
$ (for this range of masses $ \sigma_R > \sigma_L $). This can be traced back
to the fact that the main contributions to $ \sigma_L $ can be interpreted as a
heavy scalar exchange, while for $ \sigma_R $ it corresponds to a heavy vector
exchange, this leads to a factor of four in $ \sigma_R $ which is compensated
by decreasing phase space only for $ \mnu \gesim 130 \gev $.

\REF\aew{C. Arzt, \etal NSF-ITP-92-122 report;
Bulletin Board: hep-ph@xxx.lanl.gov -- 9304206 (unpublished)}
\bigskip {\bf 4.} From the above we conclude that the polarized experiment
described in this letter can be used to generate interesting bounds on the
scale of the physics underlying the standard model, even when it is assumed to
be weakly coupled. If such an effect is found the corresponding interactions
will certainly be observable in their full glory at the next colliders such as
the SSC and LHC. These bounds, though weaker than those obtained in other
experiments, such as the muon's anomalous magnetic moment [\aew], are of
interest since they probe the helicity violating processes which generate
\theops.

The sensitivity to $ \Lambda $ (understood to be the mass of a low lying
resonance) depends crucially on the estimates on the coefficients $ \lambda_i
$: if the effective operators are assumed to be generated by weakly interacting
physics, HERA's sensitivity drops by one order of magnitude compared to the
results in Refs. \polar, \prev, \blv.

Of course it is certainly possible that the operators considered here are
generated at a scale much larger than the ones of the type studied in
[\curcur], so that the non-observability of helicity violating processes does
not necessarily constrain other types of beyond the standard model effects. It
is also true that the approach followed here cannot determine the specific type
of new physics which generates the effective operators responsible for a
non-zero effect, this necessarily must await a collider that can reach the
scale $ \Lambda $.

\ack
The author gratefully acknowledges the help of M. Einhorn who also suggested
this problem.

\refout \figout

\bye